\documentclass{elsart3p}
\usepackage{amssymb}
\usepackage{graphicx}

\begin{document}

\begin{frontmatter}

\title{Fermionic systems with charge correlations on the Bethe lattice}
\author[a]{Ferdinando Mancini}
\author[a]{Adele Naddeo\corauthref{cor}}
\corauth[cor]{Corresponding author.}
\ead{naddeo@sa.infn.it}

\address[a]{Dipartimento di Fisica "E. R. Caianiello", Unit\'{a} CNISM di
Salerno\\
Universit\'{a} degli Studi di Salerno, 84081 Baronissi (SA),
Italy}

\begin{abstract}
A fermionic model, built up of $q$ species of localized Fermi
particles, interacting by charge correlations, is isomorphic to a
spin-$\frac{q }{2}$ Ising model. However, the equivalence is only
formal and the two systems exhibit a different physical behavior.
By considering a Bethe lattice with $q=1$, we have exactly solved
the models. There exists a critical temperature below which there
is a spontaneous breakdown of the particle-hole symmetry for the
first model, and of the spin symmetry for the second. While the
spin system is always stable and exhibits a homogeneous
ferromagnetic phase below $T_{c }$, the fermionic system for
$T<T_{c }$ is unstable against the formation of inhomogeneous
phases with charge separation.
\end{abstract}

\end{frontmatter}

It is known \cite{professore1} that there is an isomorphism
between fermionic models, built up of $q$ species of localized
Fermi particles, interacting by charge correlations, and
spin-$\frac{q}{2}$ Ising-like models. The fermionic system is
described by the Hamiltonian
\begin{equation}
H_{ferm}=-\mu \sum_{i}n\left( i\right)
+\frac{1}{2}zV\sum_{i}n\left( i\right) n^{\alpha }\left( i\right)
\label{1}
\end{equation}
where $n\left( i\right) =\sum_{a=1}^{q}c_{a}^{\dagger }\left(
i\right) c_{a}\left( i\right) $ is the total particle density,
$c_{a}\left( i\right) $ and $c_{a}^{\dagger }\left( i\right) $
being the annihilation and creation
operators of the species $a$ in the Heisenberg picture; $i=\left( \mathbf{i}%
,t\right) $, where $\mathbf{i}$ stands for the lattice vector
$\mathbf{R}_{i} $. These operators satisfy canonical
anti-commutation relations. $z$ is the coordination number of the
underlying lattice, $V$ is the strength of the intersite
interaction and $\mu $ is the chemical potential. The spin system
is described by the Hamiltonian
\begin{equation}
H_{spin}=-h\sum_{i}S\left( i\right) -\frac{1}{2}zJ\sum_{i}S\left(
i\right) S^{\alpha }\left( i\right)   \label{2}
\end{equation}
where $S\left( i\right) $, the spin operator at the site
$\mathbf{i}$, takes the $q+1$ values
$-\frac{q}{2},...,\frac{q}{2}$; $J$ is the exchange interaction
and $h$ is the external magnetic field. We are considering
systems with first-nearest neighbor interactions; for a generic operator $%
\Phi \left( i\right) $ we use the notation $\Phi ^{\alpha }\left(
i\right) =\sum_{j}\alpha _{ij}\Phi \left( \mathbf{j},t\right) $,
where $\alpha _{ij}$ is the projector on the first-nearest
neighbor sites. The equivalence of the two models,
$H_{ferm}=E_{0}+H_{spin}$, is based on the following relations
\begin{equation}
\begin{array}{cc}
n\left( i\right) =\frac{q}{2}+S\left( i\right)  & V=-J \\
\mu =h+\frac{q}{2}zV & E_{0}=-\frac{q}{2}\left( \mu
-\frac{q}{4}zV\right) N
\end{array}
\label{3}
\end{equation}
where $N$ is the number of sites. The relation between the
partition functions is $Z_{ferm}=e^{-\beta E_{0}}Z_{spin}$, and
the thermal average of any operator $A$ assumes the same value on
both models $\left\langle A\right\rangle _{ferm}=\left\langle
A\right\rangle _{spin}$. We have shown \cite{professore1} that
these systems are exactly solvable. This means that it is always
possible to find a complete set of eigenoperators and eigenvalues
of the Hamiltonian (\ref{1}) and/or (\ref{2}) which close the
hierarchy of the equations of motion. In such a way, formal exact
expressions for the relevant Green's functions and correlation
functions can be derived, which depend on a finite set of
parameters to be self-consistently determined. It has been shown
how to fix exactly such parameters by means of
algebra constraints \cite{professore6} in the case of a linear chain and $%
q=1,2,3$ \cite{professore1}, and in the case of the Bethe lattice
with any $z$ and $q=1$ \cite{noi}. However, the equivalence of the
two models is just formal. There is an enormous difference from a
physical point of view. The reason is the
following. For the spin system, the external thermodynamical parameters are $%
h$ and $T$: the system responses to these parameters by a certain
configuration of the spin, described by the magnetization
$m=\left\langle S\right\rangle $. For the fermionic model, the
external thermodynamical parameters are $n$ and $T$, where $n$ is
the particle density: the system responses to these parameters by
adjusting the chemical potential $\mu $. In order to illustrate
these differences, we have studied the two models (\ref
{1}) and (\ref{2}) on the Bethe lattice with $z=3,4$, by considering the case $%
J>0$ (i. e. ferromagnetic coupling). According to the exact
solution given
in Ref. \cite{noi}, there is a critical temperature $T_{c}$ such that for $%
T<T_{c}$ there is a spontaneous breakdown of the symmetry enjoyed
by the two
models: Hamiltonian (\ref{2}) is invariant under the transformation $%
S\rightarrow -S$, $h\rightarrow -h$. In the fermionic system this
transformation corresponds [cfr. (\ref{3})] to the particle-hole
transformation $\mu \rightarrow -\mu +qzV$, $n\rightarrow -n+q$. At $n=\frac{%
q}{2}$, where $\mu =\frac{zqV}{2}$, the Hamiltonian (\ref{1}) is
invariant
and enjoys the symmetry. For the spin system the critical temperature $%
T_{c}\left( h\right) $ and the magnetization $m$ are shown in
Figs. 1 and 2, respectively, for $z=3$. For general $z$,
$T_{c}\left( h\right) $ decreases from the value
$k_{B}T_{c}=\frac{2J}{\log \left( \frac{z}{z-2}\right) }$ at $h=0$
and vanishes at $\left| h\right| =J\left( z-2\right) $. As seen in
Fig. 2, for $T<T_{c}$ there is a spontaneous magnetization in zero
field.

\begin{figure}[tbph]
\centering\includegraphics*[width=0.5\linewidth]{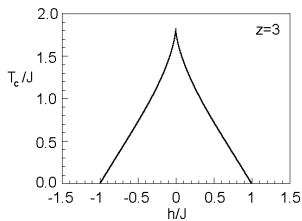}
\caption{The temperature $T_{c}\left( h\right) $ is plotted
against the magnetic field $h$ for $z=3$.} \label{figura1}
\end{figure}
\begin{figure}[tbph]
\centering\includegraphics*[width=0.5\linewidth]{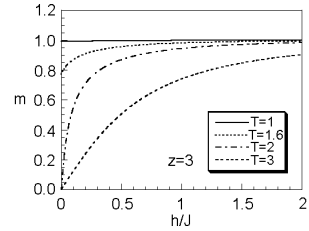}
\caption{The magnetization $m$ is plotted against $h/J$ for $z=3$
and several values of the temperature.} \label{figura2}
\end{figure}

For the fermionic system the physical situation is rather
different. The critical temperature $T_{c}\left( n\right) $ and
the chemical potential are shown in Figs. 3 and 4, respectively.
For $T<T_{c}\left( n\right) $, in correspondence of a fixed value
of the chemical potential there are three solutions for the
particle density. As clearly seen in Fig. 4, $n_{2}$ corresponds
to an unstable solution (the compressibility is negative). In
conclusion, while the spin system is always stable and exhibits a
homogeneous ferromagnetic phase below $T_{c}$, the
fermionic system for $T<T_{c}$ is unstable, except small regions around $%
n\approx 0$ and $n\approx 1$, against the formation of
inhomogeneous phases with charge separation.

\begin{figure}[tbph]
\centering\includegraphics*[width=0.5\linewidth]{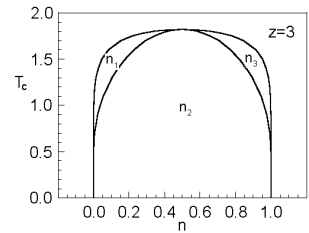}
\caption{The phase diagram in the space $T_{c}-n $ for $z=3$.}
\label{figura3}
\end{figure}
\begin{figure}[tbph]
\centering\includegraphics*[width=0.5\linewidth]{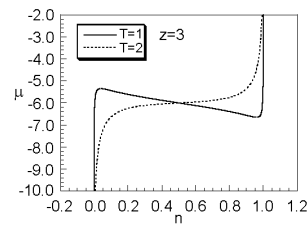}
\caption{The chemical potential $\mu$ is plotted versus $n$ for
$z=3$.} \label{figura4}
\end{figure}

\end{document}